\documentstyle[times,emulateapj]{article}
\newcommand{\PSbox}[3]{\mbox{\rule{0in}{#3}\includegraphics{#1}\hspace{#2}}}

\begin{document}
 
\title{Localizations of Thirteen Gamma-ray Bursts by the All-Sky Monitor on RXTE}

\author{D. A. Smith\altaffilmark{1}, A. M. Levine\altaffilmark{1},
H. V. Bradt\altaffilmark{1}, R. Remillard\altaffilmark{1},
J. G. Jernigan\altaffilmark{2}, K. C. Hurley\altaffilmark{2},\\
L. Wen\altaffilmark{1}, M. Briggs\altaffilmark{3},
T. Cline\altaffilmark{4}, E. Mazets\altaffilmark{5},
S. Golenetskii\altaffilmark{5}, and D. Frederics\altaffilmark{5}}

\altaffiltext{1}{Center for Space Research and Department of Physics, MIT, Cambridge, MA 02139}
\altaffiltext{2}{Space Sciences Laboratory, University of California, Berkeley, CA 94720-7450}
\altaffiltext{3}{NASA/MSFC, Code ES-84, Huntsville, AL 35812}
\altaffiltext{4}{NASA Goddard Space Flight Center, Code 661, Greenbelt, MD 20771}
\altaffiltext{5}{A. F. Ioffe Physico-Technical Institute, Politechnicheskaya 26, 194021 St. Petersburg, Russia}

\authoremail{dasmith@space.mit.edu}

\begin{abstract}

The All-Sky Monitor (ASM) on the {\it Rossi X-ray Timing Explorer}
({\it RXTE}) has been used to localize thirteen confirmed X-ray
counterparts to Gamma-Ray Bursts (GRBs) detected over three years of
operation.  We quantify the errors in ASM localizations of brief
transient sources by using observations of persistent sources with
well-known locations.  We apply the results of this analysis to obtain
accurate error boxes with reliable confidence levels for the thirteen
GRBs.  In six of these thirteen cases, multiple detections by the ASM
allow the positions to be localized to a diamond of order
$\sim15\arcmin \times 3\arcmin$.  In five further cases, the
Interplanetary Network (IPN) constrains the usually $\sim3\arcdeg
\times 3\arcmin$ (full-width) ASM error box to an area of a few tens
of square arcminutes.  This work adds eleven burst localizations to
the list of $\sim60$ well-localized GRBs.

\end{abstract}
 
\keywords{gamma rays: bursts}
 
\section{Introduction}

The recent achievement of fast, accurate localizations of Gamma-Ray
Bursts (GRBs) has led to great progress in the study of these
enigmatic events.  The Wide-Field Camera (WFC) on {\it BeppoSAX}
(Jager et al. \markcite{jag98}1998) was used to localize the X-ray
flash associated with GRB~970228 within a circle $3\arcmin$ in radius
(\markcite{cost97a}Costa et al. 1997a).  Eight hours later this circle
was imaged with the {\it BeppoSAX} Narrow-Field Instruments, and a
fading X-ray source was found and localized to within a circle of
$50\arcsec$ in radius (\markcite{cost97b}Costa et al. 1997b).  The
location of the fading source was consistent with the WFC position.
Both of these positions were widely disseminated among the
astronomical community within a few hours, and led to a cascade of
reports in the IAU Circulars (25~Circulars between Nos. 6572 and
6747), including the first report of a detection of an optical
counterpart to a GRB source (\markcite{vanpa97}van Paradijs et
al. 1997).

Since GRB~970228 was localized and identified, the WFC has been used
to provide rapid, accurate positions for almost twenty further GRBs,
and many of these positions have led to significant advances in the
understanding of GRBs.  The early successes of the {\it BeppoSAX} GRB
localization program also motivated other successful projects to use
X-ray emission to rapidly localize GRBs, such as the work reported on
in this paper and the efforts reported by Takeshima et
al. (\markcite{take98}1998).  These X-ray localizations have led to
identifications of GRB source counterparts in many wavebands, enabling
scientists to learn more about their locations, properties, and
behavior.  Optical spectroscopy of the fading counterparts of GRB
sources and what are thought to be their host galaxies has led to the
determination of five precise cosmological redshifts at the time of
this writing (perhaps six -- GRB~980425 may be a special case: see
\markcite{kulk98b}Kulkarni et al. 1998b). GRB~970508 was found to have
a redshift $z \gtrsim 0.83$ (\markcite{metzg97}Metzger et al. 1997),
the host galaxy for GRB~971214 was measured at $z = 3.4$
(\markcite{kulk98a}Kulkarni et al. 1998a), and the host for GRB~980613
shows an emission line at $z = 1.0964 \pm 0.0003$
(\markcite{djorg99}Djorgovski et al. 1999).  The counterpart to
GRB~980703 displayed emission and absorption features at $z=0.965$
(\markcite{djorg98}Djorgovski et al. 1998), and the spectrum of
GRB~990123 showed many absorption features at $z=1.600\pm0.001$
(\markcite{kelso99}Kelson et al. 1999; \markcite{hjort99}Hjorth et
al. 1999).  This is a rapidly-changing field, and new discoveries are
reported almost monthly, but these measurements indicate that at least
some GRB sources are at cosmological distances.

The observations of the temporal variation of the emission from a GRB
source in many spectral bands has made it possible to infer physical
parameters of the underlying explosive event (\markcite{wijer97}Wijers
\& Galama 1998) according to the relativistic fireball model
(\markcite{reesm92}Rees \& M\'{e}sz\'{a}ros 1992).  Further progress
in all areas of GRB investigation now depends primarily on the rapid
and accurate determination of the positions of more GRBs, so as to
quickly bring the source positions under the scrutiny of powerful
X-ray, optical, and radio telescopes.

The All-Sky Monitor (ASM) on the {\it Rossi X-ray Timing Explorer}
({\it RXTE}) continually scans the sky in the 1.5--12~keV band
(\markcite{levin96}Levine et al. 1996).  Like the {\it BeppoSAX} WFC,
it uses a proportional counter to measure the X-ray shadows of a
coded-mask, to thereby obtain fine angular resolution over a wide
field of view.  Although the design of the ASM was not optimized to
perform studies of GRB sources, on occasion it serendipitously
observes the X-ray component of a GRB. Such observations can be used
to obtain the precise position of the GRB source soon after the 
event.  The schedule of ASM observations is determined in advance and
is never adapted to respond to alerts from other instruments.

Real-time monitoring of ASM data from May 1997 to January 1999 has led
to the detection and rapid localization of six GRBs.  A search of
archival data has revealed that at least eight GRB events were
detected with the ASM in the first eighteen months of its operation,
although GRB~961216 could not be reliably localized.  Of the thirteen
GRBs localized by the ASM, only one, GRB~971214, was also detected by
the {\it BeppoSAX} WFC.  In this paper we report the celestial
positions of the ASM-detected GRB events, and we describe the
capabilities of the ASM to localize brief X-ray transients.

\setcounter{figure}{0} 
\refstepcounter{figure} 
\PSbox{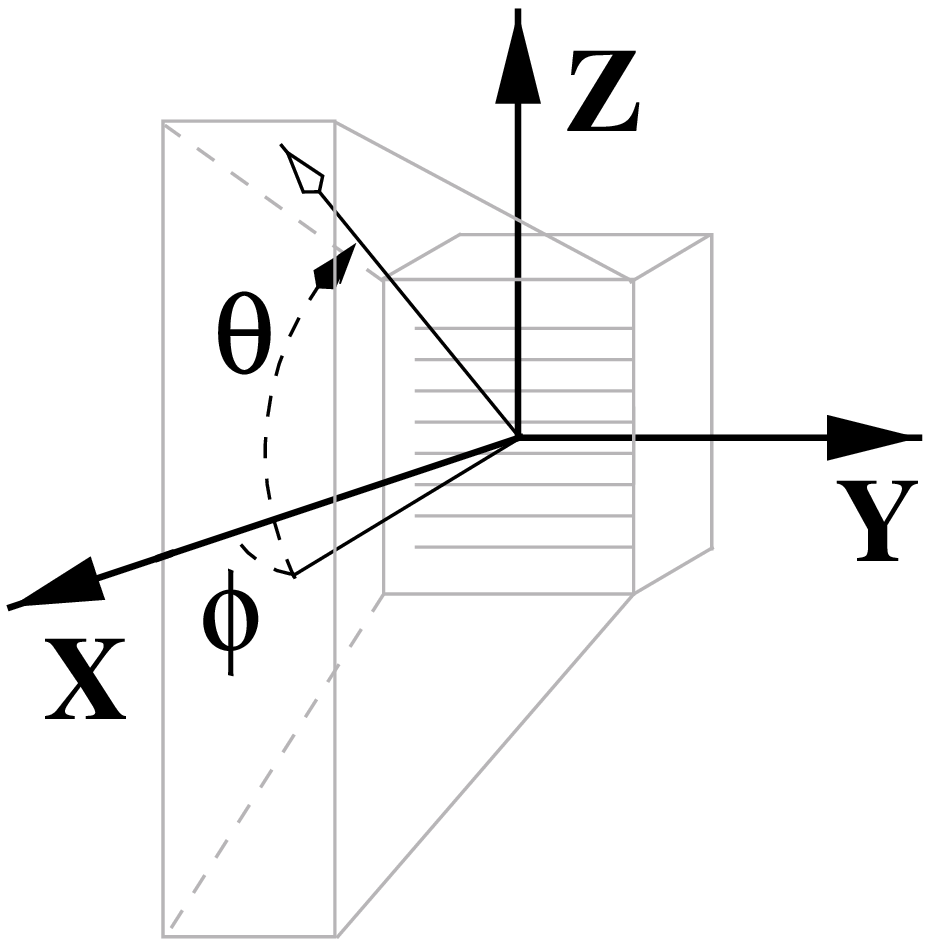 voffset=20 hscale=90 vscale=90}{3.5in}{3.7in}{\\\small 
Fig. 1 -- This diagram defines a Cartesian coordinate system and
associated angles used to describe positions and directions relative
to a particular SSC.  The coordinate axes are superposed upon a
schematic drawing of an SSC which represents the outer shell of the
collimator and the proportional counter.  The eight parallel lines
within the counter represent the resistive anodes which are used to
obtain the $y$ coordinates of detected events.  The origin of the
coordinate system is located at the center of the window of the
proportional counter.  The $x$-axis extends through the center of the
coded mask so as to point in the direction of the center of the field
of view of the SSC.  The $y$-axis points parallel to the resistive
anodes, and the $z$-axis points parallel to the long axis of the mask
slits.  A direction relative to the SSC may be specified by the angles
$\phi$ and $\theta$, where $\phi$ is measured in the $x$-$y$ plane and
$\theta$ is the angle between the $x$-$y$ plane and the specified
direction.  The field of view of the SSC extends over a region defined
by $-6\arcdeg \leq \phi \leq +6\arcdeg$ and $-55\arcdeg \leq \theta
\leq +55\arcdeg$ (FWZI).  Note that each of the three SSCs has a
unique orientation on the ASM (see Figure~1 of Levine et
al. \protect{\markcite{levin96}}1996).
\label{fovmap}}

\vspace{0.4cm}

\section{ ASM Source Position Analysis}

\subsection{Instrumentation and Analysis}

The ASM consists of three Scanning Shadow Cameras (SSCs) mounted on a
motorized rotation drive.  The assembly holding the three SSCs is
generally held stationary for a 90-s ``dwell''.  The drive then
rotates the SSCs through $6\arcdeg$ between dwells, except when it is
necessary to rewind the assembly.  Each SSC contains a proportional
counter with eight resistive anodes.  Each event detected on exactly
one resistive anode is characterized by SSC and anode numbers, total
pulse height, and a one-dimensional position in the coordinate
parallel to the anode, determined via the charge-division technique.
Each SSC views a $12\arcdeg \times 110\arcdeg$ (FWZI) field through a
mask perforated with pseudo-randomly spaced slits.  The long axes of
the slits run perpendicular to the anodes.

For each dwell, the intensities of known sources in the field of view
(FOV) are derived via a fit of model slit-mask shadow patterns to
histograms of counts as a function of position in the detector.  The
residuals from this fit are then cross-correlated with each of the
expected shadow patterns corresponding to one of a set of possible
source directions which make up a grid covering the FOV.  A peak in
the resulting cross-correlation map indicates the possible presence
and approximate location of a new, uncataloged X-ray source.  Peaks
that satisfy certain criteria are analyzed further to confirm the
detection and refine the position.  For detections of bright sources,
the resulting error boxes typically have sizes on the order of
$3\arcmin \times 3\arcdeg$ full width at 90\% confidence.  Detections
of new sources in at least two SSCs are preferred because they yield
error boxes that cross, thereby constraining the source's location in
two dimensions to a diamond-shaped error box, of dimensions
$\sim3\arcmin \times 15\arcmin$ for bright sources.

Errors in the model of the shadow patterns from known sources in the
FOV can lead to ``ripples'' in the cross-correlation map which act as
a source of noise in addition to counting statistics.  The correct
construction of the shadow patterns depends on an accurate calibration
of the correspondence between electrical position, which is directly
derived via application of the charge-division technique, and physical
location within the detector.  The electrical position - physical
location correspondence in the SSCs is drifting with time as the
carbon coating of the resistive anodes is being worn away
nonuniformly.  We therefore periodically update our calibration of
this correspondence.

\subsection{The Error Distributions}

\begin{deluxetable}{ccccccc}
\tablecaption{Calibration Sources \label{sotab}}
\tablehead{
Source Name &
R.A.  &
Decl. &
Intensity &
No. in &
No. in &
No. in \nl
 &
(J2000) &
(J2000) &
(mCrab)\tablenotemark{a} &
SSC 1\tablenotemark{b} &
SSC 2\tablenotemark{b} &
SSC 3\tablenotemark{b} 
}
\startdata
Sco X-1 & 16h 19m 55.13s &  $-15\arcdeg38\arcmin24.4\arcsec$ & 9000-23000 & 755 & 816 & 1153 \nl
Crab & 05h 34m 31.97s & $+22\arcdeg00\arcmin52.2\arcsec$ & 930-1100\tablenotemark{c} & 598 & 546 & 1644 \nl
Cyg X-2 & 21h 44m 40.97s & $+38\arcdeg19\arcmin18.1\arcsec$ & 200-900 & 1148 & 1151 & 353 \nl 
Cyg X-3 & 20h 32m 25.54s & $+40\arcdeg57\arcmin27.7\arcsec$ & 70-600 & 1378 & 1368 & 400 \nl
Her X-1 & 16h 57m 49.73s & $+35\arcdeg20\arcmin32.3\arcsec$ & $<200$ & 222 & 218 & 44 \nl
X 0614+091 & 06h 17m  7.32s & $+09\arcdeg08\arcmin13.6\arcsec$ & $<200$ & 416 & 293 & 300 \nl
\enddata
\tablenotetext{a}{Intensity in mCrab derived from detected count
rates, adjusted to simulate observations with the source at the center
of the FOV of SSC 1, and normalized to a nominal count rate of 75 c/s
for the Crab Nebula at the center of the FOV of SSC 1.}
\tablenotetext{b}{The number of observations of each source used
to analyze the position error.}
\tablenotetext{c}{Including statistical and systematic errors in
ASM detections.}
\end{deluxetable}

For a GRB source position, it is essential to report an accurate error
box with a reliable confidence level, i.e., a good estimate of the
probability that the source is within the box.  In this section, we
describe our method of determining the association between error box
size and confidence level.  The method is empirical and relies on
observations of persistent X-ray sources.  If a source is removed from
the catalog of known sources used in the initial fit, it functions as
a new source to be detected and localized with our standard software.
The derived location may then be compared with the known precise
location of that source's optical counterpart.

\begin{figure}
\plotone{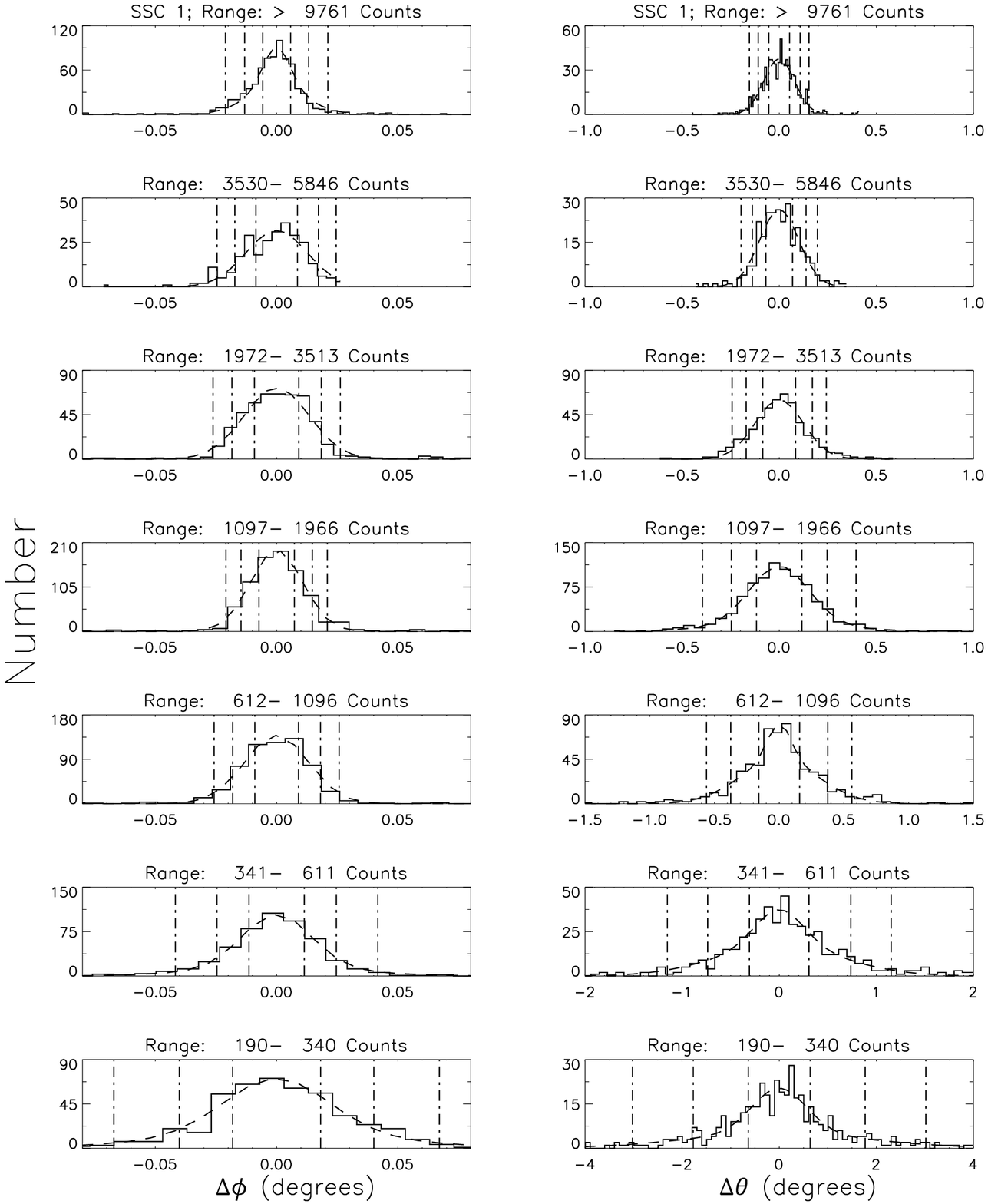}
\caption{A demonstration of the position-determining ability of SSC~1
and the validity of the model used to calculate confidence limits.
Each histogram shows the distribution of errors in $\phi$ (left
panels) and $\theta$ (right panels) between the derived source
position and the actual celestial position for detections of sources
with measured source counts within a certain range.  The number of
counts increases vertically in the figure.  Note that the abscissa
scale is not the same in each panel.  The dashed curve represents the
best-fit model of two summed Gaussian curves and a constant term to
these data.  The presence of the constant term is not obvious, because
the abscissa scale is defined to emphasize the central peak. Broken
vertical lines indicate the symmetric zones containing 50\%, 82\% and
95\% of the area under the double-Gaussian curve.}
\label{er0}
\end{figure}

This procedure was performed on 13,982 observations of six X-ray
sources, chosen to provide a wide range of intensities
(Table~\ref{sotab}).  We chose the test observations from five 50-d
time intervals when the six test sources were observed most often.
The best-fit position from each observation of one of the six sources
was compared with the catalogued source position to obtain deviations
in both FOV coordinates, $\phi$ and $\theta$ (Fig.~\ref{fovmap}).  The
distribution of the resulting position errors provides the basis for
the association between confidence levels and error box sizes
(Fig.~\ref{er0}).

There are four main characteristics of an observation that can affect
the accuracy of attempts to localize a source within that observation:
the source's intrinsic brightness, the source's FOV location, the
contribution to the background from other sources in the FOV, and the
number of diffuse background counts.  One advantage to the coded-mask
imaging technique is that the background strength to which the signal
strength should be compared is not the total number of non-signal 
counts in the

\end{multicols}
\section*{}
\vskip -15pt

\noindent
entire detector, but rather the number of non-signal counts detected
in that portion of the detector which is illuminated by the (off-axis)
source of interest.

The dominant contribution to the total background is usually from
diffuse celestial X-ray emission for which $\sim2500$ counts are
typically detected per observation.  A relatively small number of
counts, $\sim$150--450, are contributed from non-X-ray background.
Occasionally, substantial fluxes of solar X-rays scattered in the
Earth's atmosphere or the inside surface of the one of the collimators
are detected.  This solar X-ray contamination can contribute up to
$\sim1600$~counts to the total background.  For the $\sim$14,000
observations used in this analysis, the median number of counts in a
single SSC per dwell (excluding the counts from the test source) is
$\sim3150$ (1.5--12~keV).  Only $\sim20$\% of the observations had a
total background higher than 5000~counts, and only $\sim5$\% of the
observations had more than 9750~background counts.  Half of the
observations show a total background between 2000~and 4000~counts,
indicating that there were no other strong discrete sources in the FOV
besides the test source.

Observations that have more than about 4000~total background counts
most often have other point X-ray sources in the FOV.  These other
sources can increase the statistical noise in an observation, as well
as the likelihood for systematic errors in a derived position.
However, the effect of a widely separated source on the derived
position for the test source tends to be small, since the overlap
between the shadow patterns of such sources is generally small.  To
minimize the effect of other sources on our results, the six test
sources were selected in part because they are relatively isolated on
the sky.  Only two of these six sources are within $15\arcdeg$ of
another X-ray source bright enough to be detected by the ASM.  The
smallest separation is the $\sim9\arcdeg$ separation of Cyg~X-3 from
Cyg~X-1.  In the case of Sco~X-1, nearby Galactic plane sources
contribute significantly to the total background, and indeed, Sco~X-1
is the test source in 73\% of the 1271 observations that have more
than 8000 total background counts.  However, Sco~X-1 is so bright (an
on-axis, 90-s observation typically yields more than 68,000~counts)
that contamination by these sources is not expected to have any
significant effect on the localization accuracy.  We therefore did not
expect, {\it a priori}, that the presence of other sources in the FOV
would often have a large effect upon the local statistical noise at
the positions of the test sources.

If the counts from other sources are not properly modeled and
subtracted via the initial fit, the noise level near the test source
may be enhanced by systematic ``ripples'' in the cross-correlation
map, even if the sources are well-separated in the FOV.  To minimize
these ripples, the calibrated relations between electrical position
and physical location at a time near the middle of each interval were
used in the generation of the model shadow patterns used to fit
observations taken during that interval.  The use of {\it a posterori}
calibration ensures that the detector response will be modeled as
accurately as possible, which means not only that the test sources
will be localized as accurately as possible, but that other discrete
X-ray sources will have a minimal effect on systematic noise near the
source of interest.  For real-time operations, we generate model
shadow patterns based on an extrapolation of the two most recent
calibration measurements.  Over months, the true calibration will
drift from this set of extrapolated values, degrading the localization
accuracy and increasing the systematic noise.  We therefore update the
calibration periodically, but there may be times when the localization
accuracy in real-time operations will not achieve the level presented
here.

We parameterize the position error as a function of the total number
of counts observed from a source during a single observation.  The
collimator reduces the area of the detector exposed to an off-axis
source, so that sources of the same intrinsic brightness will yield
different numbers of counts if they are observed at different
locations in the FOV.  For example, a $\sim300$ mCrab source at the
center of the FOV will generate $\sim2000$~counts in a 90-s
observation, while the same source at $\phi=\pm4.0\arcdeg$ will only
contribute $\sim700$ counts.  The FOV position of the relevant test
source is effectively random.  Thus, the 14,000 test observations
yield numbers of source counts that span the entire range expected
from a single source during a 

\refstepcounter{figure} 
\PSbox{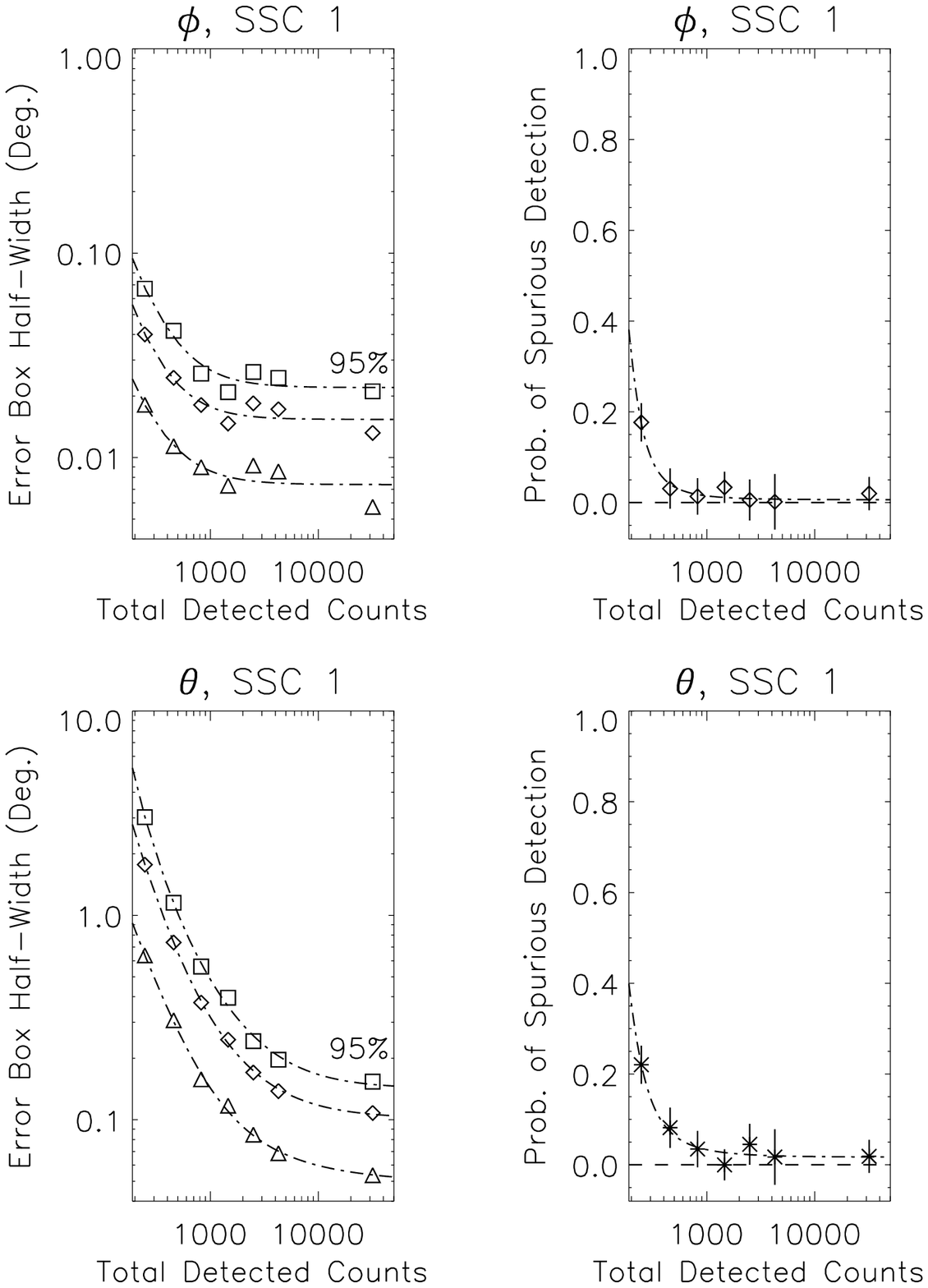 hoffset=-25 voffset=0 hscale=64 vscale=62}{3.5in}{5.0in}
{\\\small Fig. 3 -- The functions used to set the size of the ASM
error box for a new X-ray source observed by SSC~1.  The left panels
display the half-widths of error boxes in each of two dimensions at
three levels of confidence (95\%, 82\%, and 50\%), as modeled by the
two-Gaussian model for the histograms shown in
Figure~\protect{\ref{er0}}.  The right panels display the fraction of
the measurements unaccounted for by that model, which we assume to be
noise peaks mistaken for sources.  All graphs use the total measured
counts from the source as the abscissa.\label{in0}}

\vspace{0.4cm}

\noindent
90-s ASM dwell.

For this analysis, we assume that an off-axis bright source and an
on-axis dim source that both yield the same number of counts in an
observation can be localized with approximately the same accuracy.
The off-axis source illuminates the detector through fewer mask slits,
yielding fewer shadow edges in the position histograms that can be
used to localize the source in the cross-correlation process.
However, each slit edge will be defined by a greater count rate
contrast, and the effective number of background counts will be
reduced in proportion to the reduction in the exposed area of the
detector.  If the localization accuracy were to scale like the
detection sensitivity, we would expect it to be approximately
inversely proportional to the square root of the exposed fraction of
the FOV, if the numbers of source and background counts are held
constant.  Therefore, over a restricted region of the FOV, any error
in estimating error box sizes using the number of source counts rather
than a more sophisticated estimator is expected to be limited.  For
sources near the edge of the FOV, this approximation is poor and
calibration errors have a particularly large effect, so we only
consider observations where the test source was located such that
$\mid\phi\mid<4.6\arcdeg$ or $\mid\theta\mid<45\arcdeg$.

We checked whether the effects of the test sources' FOV locations are
adequately approximated by our assumption that the localization
accuracy depends only on the number of source counts.  To do this, we
separated the $\sim14,000$ observations into two groups according to
the $\phi$ coordinate of the test source's actual location, i.e., the
two groups were defined by $\mid\phi\mid<2.0\arcdeg$ and $2.0\arcdeg
\leq \mid\phi\mid \leq 4.6\arcdeg$.  We binned the deviations between
the actual and derived source locations according to number of source
counts and FOV group and obtained two sets of error histograms.  The
widths of the resulting error distributions for either the $\theta$ or
$\phi$ coordinate did not show any clear systematic dependence on FOV
group.

The effect of the background level on the accuracy of our
localizations was similarly checked by dividing the set of
observations into groups based on the total background, i.e., all
counts excluding those from the test source.  We again calculated
position error distributions based on test source counts for each of
these groups.  The error distributions for the high background cases
($>4000$~counts) were found to be indistinguishable within statistical
limits from the distributions for the low background cases
($<4000$~counts).  The cases in which the number of total background
counts rose above 8000 were also examined.  As noted, 73\% of these
observations used Sco X-1 as the test source.  No significant
difference between the error distributions from these observations
with high total background and those of low total background was
found.  These results indicate that sources widely separated from the
test source have only small effects on the localization of the test
source.  A new source that appears close to a strong source will not
be localized as accurately as indicated by the present results, but we
have not tried to quantify the effects of other nearby sources.

We therefore proceed as follows: for each of the $\sim14,000$
observations we record both the total number of photons detected from
the test source and the deviation of the derived position in $\phi$
and $\theta$ from the true position.  We separate the location
measurements according to the number of detected source counts into
seven groups per SSC and bin the angular errors for all observations
in each group into histograms.  Figure~\ref{er0} shows the resulting
distributions for SSC~1.  The equivalent histograms for the other two
SSCs look similar.

\subsection{A Model for the Error Boxes\label{sec:mod}}

We modeled each histogram of coordinate errors as the sum of a narrow
Gaussian, a wide Gaussian, and a constant.  For bright sources, a
single narrow Gaussian is sufficient to obtain a reasonable fit.  For
small values of source counts, there is a significant chance that
noise in the cross-correlation map will result in an incorrect
identification of the source location.  The noise may raise a side
lobe of the instrument response to a value higher than the peak at the
source location.  This yields broad wings on the side of the central
peak in the error histograms.  These wings are modeled by the wide
Gaussian.  Noise peaks may even exceed the value of any response to
the test source, which results in a population of measured errors that
are distributed uniformly across the whole range of allowed values for
each angle.  This uniform scatter is present but not obvious in
Figure~\ref{er0}, because only the region near the central peak is
displayed.  At extremely low source counts, the central Gaussian 
peak disappears entirely, leaving only the random scatter.  There is
no evidence for a central peak below $\sim200$~counts in SSCs~1 and 2
and below $\sim300$ counts in SSC~3.

\refstepcounter{figure} 
\PSbox{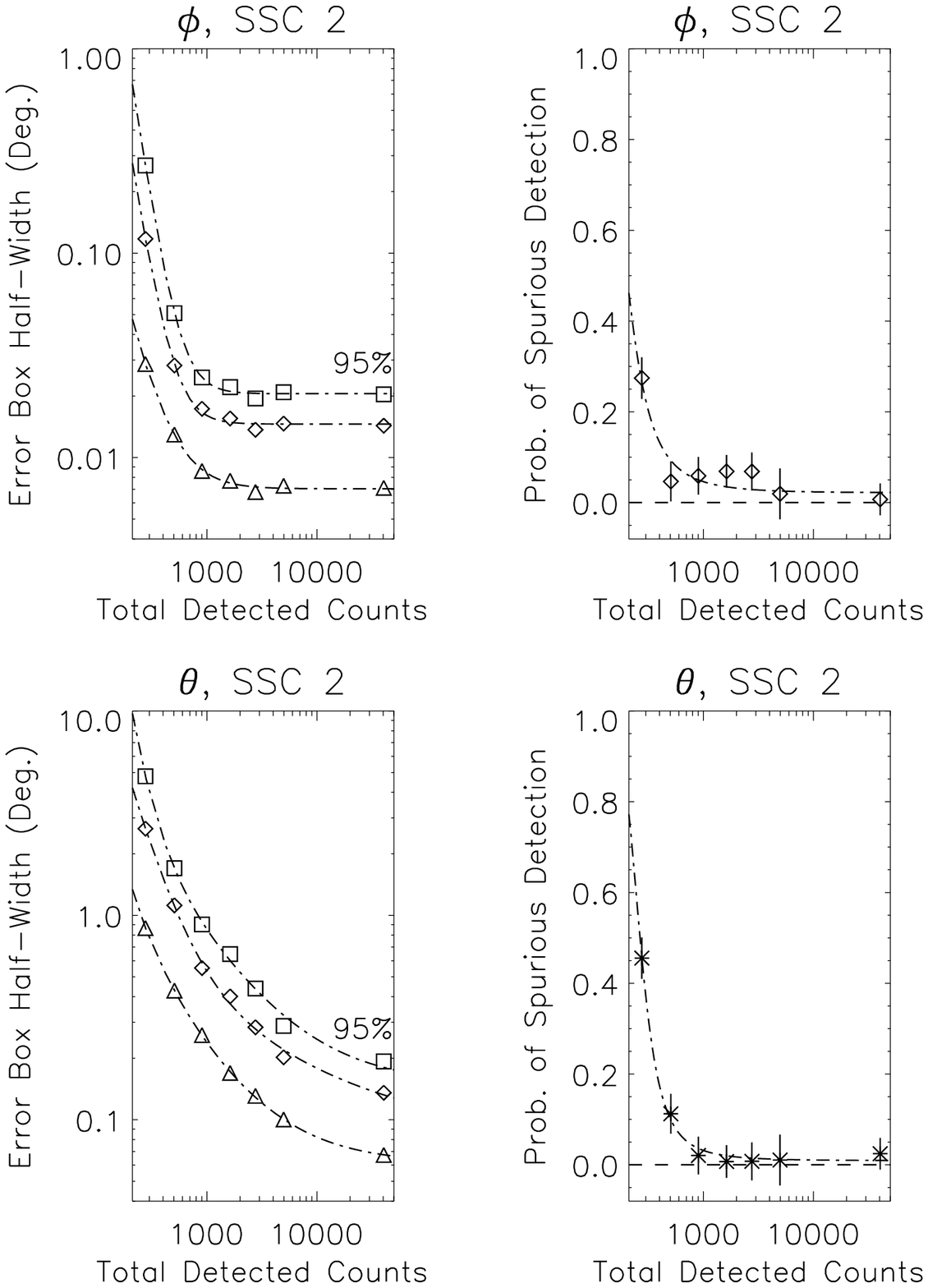 hoffset=-25 voffset=0 hscale=64 vscale=62}{3.5in}{5.0in}
{\\\small Fig. 4 -- The functions used to set the size of the ASM
error box for a new X-ray source observed by SSC~2.  See caption for
Figure~\protect{\ref{in0}}.\label{in1}}

\vspace{0.4cm}

Our strategy in reporting error boxes is to define uncertainties 
in each FOV coordinate appropriate to a given confidence level under
the assumption that the detection was of a real source.  In addition,
we quote the probability that the detection was spurious.  The
estimated probabilities are derived by integrating under the
components of the best-fit model distribution.  A confidence interval
at $x$\% is defined as the symmetric interval about zero such that the
integral of the two Gaussian curves over that interval yields $x$\% of
the integral of the same Gaussians over the entire range of possible
errors.  Figure~\ref{er0} displays as vertical broken lines the values
in both the $\phi$ and $\theta$ directions that correspond to 50\%,
82\% and 95\% confidence limits for SSC~1.

The uncertainty in each coordinate associated with each of the
selected confidence limits is plotted as a function of the number of
detected source counts in the left panels of
Figures~\ref{in0}--\ref{in2}.  We used these results to derive
interpolation functions which give FOV coordinate uncertainties for
the source location at three significance levels for any number of
source counts.  We chose as an interpolation function the sum of two
power laws and a constant.  The constant represents the limiting
systematic error (with values of $\sim1.5\arcmin$ in $\phi$ and
$12\arcmin$ in $\theta$ at 95\% confidence), and the power laws are
simply a convenient means to interpolate between the estimated angular
uncertainties.  The best-fit interpolation functions are graphed as
broken lines in the left panels of Figures~\ref{in0}-\ref{in2}.

The difference between the total number of measurements in the
histogram and the area under the two model Gaussian 

\refstepcounter{figure} 
\PSbox{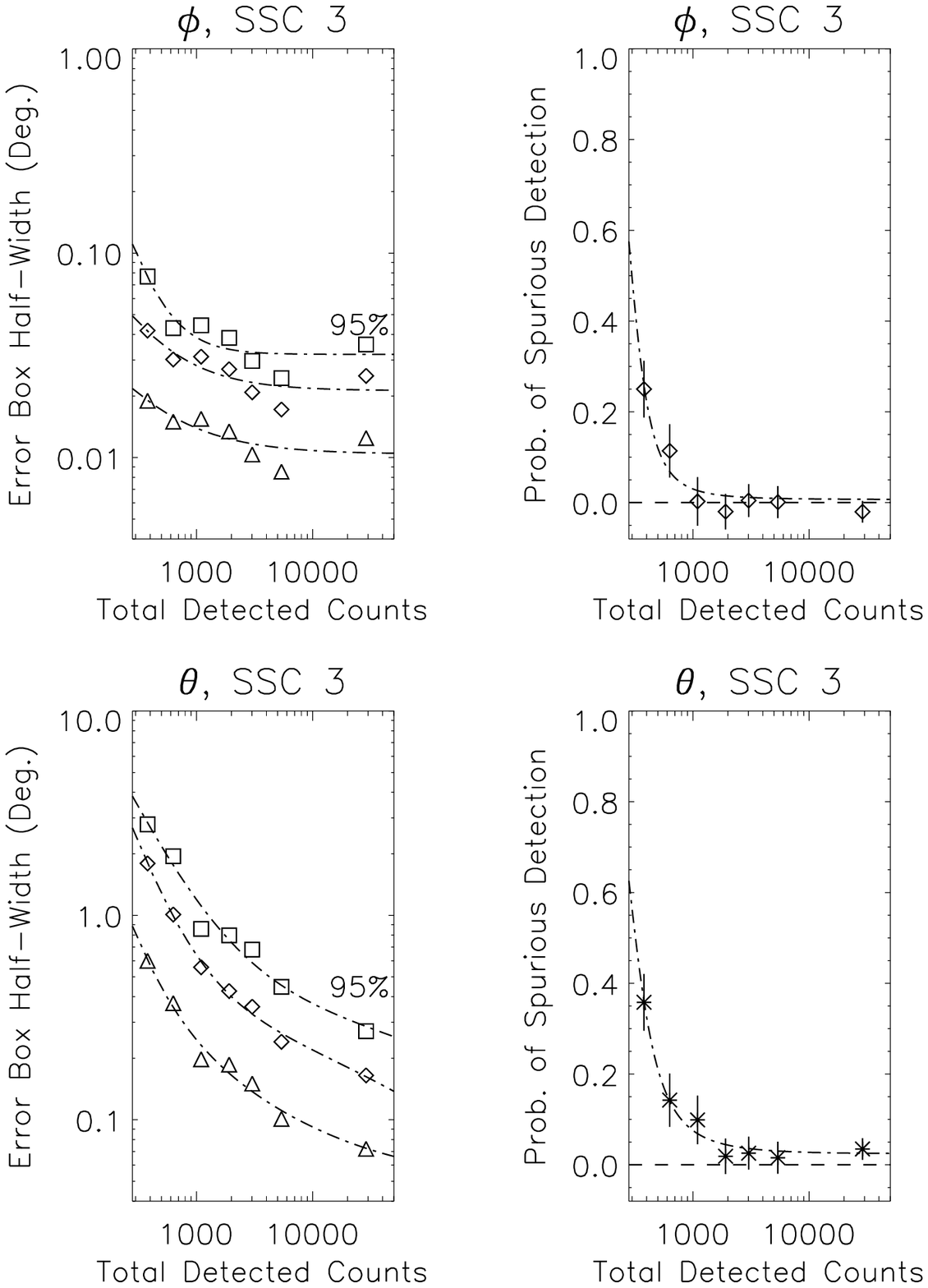 hoffset=-25 voffset=0 hscale=64 vscale=62}{3.5in}{5.0in}
{\\\small Fig. 5 -- The functions used to set the size of the ASM
error box for a new X-ray source observed by SSC~3.  See caption for
Figure~\protect{\ref{in0}}.\label{in2}}

\vspace{0.4cm}

\noindent
curves is a measurement of the probability of mistaking a noise peak
for the source.  This difference is plotted in the right panels of
Figures~\ref{in0}-\ref{in2}. The error bars reflect the counting
statistics of the total number of actual measurements in each
histogram, and are typically 3--7\%.  These figures show that the
probability of misidentifying a source increases as the number of
detected photons decreases.  To interpolate between these points, we
use the formula:
\begin{equation}
\label{spur}
f(x) = \frac{1.0}{1 + e^{-b/(x - a)}}
\end{equation}
where $x$ is the number of detected source counts, and $a$ and $b$ are
constant parameters determined by a least-squares fitting procedure.
We chose this function because it could match the rising behavior of
the data at low numbers of counts, while also ensuring that the
probability never exceeded one or dropped below zero.  The best-fit
functions are plotted as broken lines in the right panels of
Figures~\ref{in0}--\ref{in2}.  Note that extrapolation of this
function below the 200--300 count limit mentioned above is not
meaningful.

\begin{deluxetable}{cccrcc}
\tablecaption{Properties of 13 ASM-detected GRBs \label{flutab}}
\tablehead{
Date of GRB &
Time of GRB &
Confirming &
2-12 keV Fluence & 
R.A. &
Decl. \nl
(yymmdd) & 
(hh:mm:ss) &
Satellite\tablenotemark{a}  &
($10^{-7}$ ergs cm$^{-2}$) & 
(J2000) &
(J2000) 
}

\startdata
960416 & 04:58:59 & ub   & $  6.0\pm0.3$ & 04h15m27s & $+77\arcdeg10\arcmin$ \nl
960529 & 05:34:34 &  k   & $>17.5\pm0.6$ & 02h21m50s & $+83\arcdeg24\arcmin$ \nl
960727 & 11:57:36 & uk   & $  9.5\pm0.5$ & 03h36m36s & $+27\arcdeg26\arcmin$ \nl
961002 & 20:53:55 & uk   & $  9.2\pm0.5$ & 05h34m46s & $-16\arcdeg44\arcmin$ \nl
961019 & 21:08:11 & ub   & $  4.6\pm0.6$ & 22h49m00s & $-80\arcdeg08\arcmin$ \nl
961029 & 19:05:10 & k    & $  3.3\pm0.4$ & 06h29m27s & $-41\arcdeg32\arcmin$ \nl
961230 & 02:04:52 & u    & $  1.5\pm0.3$ & 20h36m45s & $-69\arcdeg06\arcmin$ \nl
970815 & 12:07:04 & ubks & $>33.3\pm0.8$ & 16h08m33s & $+81\arcdeg30\arcmin$ \nl
970828 & 17:44:37 & ub   & $>14.9\pm0.6$ & 18h08m23s & $+59\arcdeg19\arcmin$ \nl
971024 & 11:33:32 & b    & $  1.1\pm0.3$ & 18h25m00s & $+49\arcdeg27\arcmin$ \nl
971214 & 23:20:41 & ubks & $  3.4\pm0.3$ & 12h04m56s & $+64\arcdeg43\arcmin$ \nl
980703 & 04:22:45 & ub   & $>18.3\pm0.8$ & 23h59m04s & $+08\arcdeg33\arcmin$ \nl
981220 & 21:52:21 & uks  & $ 12.6\pm0.5$ & 03h43m38m & $+17\arcdeg13\arcmin$ \nl
\enddata
\tablenotetext{a}{u - {\it Ulysses}; b - BATSE; k - KONUS; s - {\it BeppoSAX}}
\end{deluxetable}

It is our practice to define error box sizes based on the
double-Gaussian integration limits (left panels;
Figs.~\ref{in0}--\ref{in2}) and to separately specify the probability
that the detection is spurious (right panels;
Figs.~\ref{in0}--\ref{in2}). The former yield a probability $p$
($=$~0.95, 0.82 or 0.50) that the actual GRB location lies within the
associated interval.  There is a probability of $p^2$ that the actual
GRB location lies within two of these intervals.  A box on the sky,
defined by 95\% limits in each of two directions, is therefore a 90\%
confidence region.  In the case of a single-SSC detection, the two
intervals will be in terms of $\phi$ and $\theta$ in that SSC, but if
detections of a given source are available in two SSCs, the diamond
formed by the intersection of the two $\phi$ intervals can be taken as
the 90\% region (See Figure~\ref{tgrb}).  Similarly, using the 82\%
values in each direction will yield an error box at 68\% confidence,
and the 50\% values will yield a joint error box at 25\% confidence.
In cases of weak source detections, this smallest box helps illustrate
the deviations of the actual error distributions from simple
Gaussians.

We have thus derived functions to estimate source position
uncertainties corresponding to the 95\%, 82\%, and 50\% confidence
limits in each of two dimensions, valid for detections of more than
200-300 counts under the assumption that the corresponding peak in the
cross-correlation map represents a valid source detection.  We find
that the width of the error distribution can be effectively
parameterized as a function of the total number of counts detected
from a source.  This analysis is valid for the central region of the
FOV, when the sky around the new source is free of contaminating
sources, and an accurate calibration of the electrical-physical
position correspondence in the detector is available.  Under these
conditions (which may not hold for all real-time detections), this
analysis provides a rational means for associating error-box sizes
with confidence levels.

\section{Results}

In this paper, we report the positions of thirteen Gamma-Ray Bursts
(GRBs) localized with the ASM in its first three years of operation.
It is impossible to unambiguously distinguish a GRB from an X-ray
burst on the basis of ASM data alone.  We therefore report the
positions of thirteen events which have been identified as GRBs
through the comparison of ASM timing and location information with
data from other operational GRB detectors including BATSE on the {\it
Compton Gamma Ray Observatory}, the Gamma-Ray Burst Instrument on {\it
Ulysses}, KONUS on the {\it Wind} spacecraft, and the Gamma-Ray Burst
Monitor on {\it BeppoSAX} (See Table~\ref{flutab}). These instruments,
as well as the Interplanetary Network (IPN) triangulation technique,
are described elsewhere (See, e.g. \markcite{fishm95}Fishman et
al. 1995; \markcite{hurl99a,hurl99b}Hurley et al. 1999a, b;
\markcite{brig99}Briggs et al. 1999; the IPN web site
http://ssl.berkeley.edu/ipn3/).

Pertinent data on these thirteen bursts are given in
Tables~\ref{flutab}--\ref{ipntab}.  Table~\ref{flutab} gives the onset
times, 1.5--12~keV fluences, and centers of the ASM error boxes,
Table~\ref{boxtab} gives the error box sizes and position angles for
the single-SSC detections, and Table~\ref{cortab} gives the celestial
coordinates of the corners of the best intersection diamonds in cases
of two-SSC detections.  Table~\ref{othtab} reports the circular
localizations achieved with other high-energy instruments, while
Table~\ref{ipntab} describes the available IPN annuli.

\begin{deluxetable}{crrr}
\tablecaption{Properties of Single-SSC ASM Error Boxes\label{boxtab}}
\tablehead{
Date of GRB &
Length &
Width &
Position Ang.\tablenotemark{a} \nl
(yymmdd) & 
(arcmin.) & 
(arcmin.) &
(deg.) 
}

\startdata
960727 &  65.4 &  2.5 &  51.85 \nl
961002 &  76.3 &  2.6 &  82.19 \nl
961019 & 493.2 & 25.6 &  52.34 \nl
961029 & 522.5 & 28.0 & -54.71 \nl
971024 & 249.5 &  6.5 &  64.38 \nl
971214 & 204.2 &  5.7 & -62.54 \nl
981220 &  58.9 &  2.5 &  75.64 \nl
\enddata
\tablenotetext{a}{Defined such that positive is east of north.}
\end{deluxetable}

\begin{deluxetable}{ccccc}
\tablecaption{Corners of Multiple-SSC Error Boxes\label{cortab}}
\tablehead{
Date of GRB &
R.A. 1 &
R.A. 2 &
R.A. 3 &
R.A. 4 \nl
(yymmdd) &
Decl. 1 &
Decl. 2 &
Decl. 3 &
Decl. 4 \nl
        &
(J2000) &
(J2000) &
(J2000) &
(J2000) \nl
}
 
\startdata
960416 & 04h15m32.8s & 04h13m30.0s & 04h15m21.9s & 04h17m25.4s \nl   
       & $+77\arcdeg11\arcmin17\arcsec$ & $+77\arcdeg07\arcmin20\arcsec$ & $+77\arcdeg08\arcmin00\arcsec$ & $+77\arcdeg11\arcmin54\arcsec$ \nl   
960529 & 02h21m19.8s & 02h19m57.3s & 02h22m17.3s & 02h23m43.3s \nl   
       & $+83\arcdeg25\arcmin33\arcsec$ & $+83\arcdeg18\arcmin07\arcsec$ & $+83\arcdeg22\arcmin38\arcsec$ & $+83\arcdeg30\arcmin03\arcsec$ \nl
961230 & 20h37m25.6s & 20h36m34.9s & 20h35m58.9s & 20h36m50.3s \nl   
       & $-68\arcdeg58\arcmin27\arcsec$ & $-68\arcdeg44\arcmin07\arcsec$ & $-69\arcdeg13\arcmin09\arcsec$ & $-69\arcdeg27\arcmin29\arcsec$ \nl
970815 & 16h08m18.7s & 16h06m16.7s & 16h08m47.1s & 16h10m51.1s \nl   
       & $+81\arcdeg32\arcmin00\arcsec$ & $+81\arcdeg27\arcmin14\arcsec$ & $+81\arcdeg29\arcmin32\arcsec$ & $+81\arcdeg34\arcmin15\arcsec$ \nl
970828 & 18h08m42.4s & 18h07m58.6s & 18h08m36.9s & 18h09m20.7s \nl   
       & $+59\arcdeg19\arcmin33\arcsec$ & $+59\arcdeg18\arcmin08\arcsec$ & $+59\arcdeg17\arcmin00\arcsec$ & $+59\arcdeg18\arcmin25\arcsec$ \nl
980703 & 23h59m07.0s & 23h59m23.3s & 23h59m02.0s & 23h58m45.7s \nl   
       & $+08\arcdeg32\arcmin05\arcsec$ & $+08\arcdeg38\arcmin01\arcsec$ & $+08\arcdeg34\arcmin43\arcsec$ & $+08\arcdeg28\arcmin47\arcsec$ \nl
\enddata
\end{deluxetable}

\begin{deluxetable}{ccccc}
\tablecaption{Sizes of Error Circles\label{othtab}}
\tablehead{
Date of GRB &
Instrument &
R.A. of Center &
Decl. of Center &
Radius \nl
(yymmdd) &
 &
(J2000) &
(J2000) &
(Degrees) 
}

\startdata
960416 & BATSE & 04h27m24s & $+73\arcdeg36\arcmin 0\arcsec$ & 4.8\nl
961019 & BATSE & 21h27m29s & $-84\arcdeg10\arcmin50\arcsec$ & 5.0\nl
970815 & BATSE & 15h37m58s & $+81\arcdeg42\arcmin 0\arcsec$ & 4.0\nl
970815 & {\it ASCA}  & 16h06m54s & $+81\arcdeg30\arcmin34\arcsec$ & 0.0167\tablenotemark{a} \nl
970828 & BATSE & 17h56m53s & $+59\arcdeg25\arcmin10\arcsec$ & 4.8\nl
970828 & {\it ASCA} & 18h08m30s & $+59\arcdeg19\arcmin20\arcsec$ & 0.025\tablenotemark{b} \nl
970828 & {\it ROSAT} & 18h08m31.7s & $+59\arcdeg18\arcmin50\arcsec$ & 0.00278\tablenotemark{c} \nl
971024 & BATSE & 18h01m53s & $+49\arcdeg38\arcmin20\arcsec$ & 6.9\nl
971214 & BATSE & 12h03m19s & $+66\arcdeg12\arcmin40\arcsec$ & 4.1\nl
971214 & {\it BeppoSAX} - WFC & 11h56m30s & $+65\arcdeg12.0\arcmin$ & 0.065\tablenotemark{d} \nl
971214 & {\it BeppoSAX} - NFI & 11h56m25s & $+65\arcdeg13\arcmin11\arcsec$ & 0.0167\tablenotemark{e}\nl
980703 & BATSE & 23h56m17s & $+12\arcdeg00\arcmin40\arcsec$ & 4.0\nl
980703 & {\it BeppoSAX} - NFI & 23h59m07s & $+08\arcdeg35\arcmin33\arcsec$ & 0.0139\tablenotemark{f}\nl
\enddata
\tablerefs{a - Murakami et al. 1997a; b - Murakami et al. 1997b; c - Greiner et al. 1997; d - Heise et al. 1997; e - Antonelli et al. 1997; f - Galama et al. 1998}
\end{deluxetable}

The GRBs described in this paper were discovered through the use of
four different techniques.  First, we searched the ASM data at the
approximate times and rough locations (the radii of the error circles
were not available at the time of our search) of 438 GRBs determined
from BATSE data between 1996 Jan~6 and 1997 Jul~15 (see, e.g.,
http://www.batse.msfc.nasa.gov/data/grb/catalog/ for times and
locations). If the trigger time of a burst corresponded to a time when
the ASM was collecting data, and if the BATSE localization was within
the ASM FOV at that time, the relevant ASM observation was searched
for evidence of a GRB detection.  This search resulted in the
discovery of X-ray counterparts for GRB~960416 (Figure~\ref{tgrb}a)
and GRB~961019 (Figure~\ref{tgrb}e).  The former event was detected by
two SSCs simultaneously, while the latter was only detected by SSC~2.
Both events were also detected by the {\it Ulysses} GRB instrument, so
IPN annuli could be calculated.  Both annuli were consistent with the
ASM positions, and in the case of GRB~961019, the IPN annulus reduced
the length of the error box to $11\arcmin$.

Second, we searched through the ASM time-series data up to September
1997 for episodes of transient emission that could be from GRBs not
detected by BATSE.  We performed linear least-squares fits to the
count rates for each 90-s dwell of the ASM.  Steady or very slowly
changing count rates (on time-scales between $1/8$ and 90 s) yielded
low values of $\chi^2$.  We then examined the observations yielding
the highest values of $\chi^2$ for GRB-like events.  We excluded cases
where a bright, persistent source moved in or out of Earth occultation
or a known highly variable source like GRS~1915+105 was in the FOV.
We searched the position data from the remaining dwells for
indications of new sources.  Observations containing a new source
candidate with an intensity measurement of 5~$\sigma$ significance or
greater in addition to excess short-term variability in the
time-series data were flagged as containing possible GRB candidates.
This search yielded five additional events that could be confirmed as
GRBs by {\it Ulysses} and/or KONUS detections.  There were no other
obvious GRB-like events that we could not identify as coming from
previously known X-ray sources.

\begin{deluxetable}{cccccc}
\tablecaption{Dimensions of IPN Annuli\label{ipntab}}
\tablehead{
Date of GRB &
Instruments\tablenotemark{a}&
R.A. of Center &
Decl. of Center &
$3-\sigma$ Full-width &
Radius \nl
(yymmdd) &
 &
(J2000) &
(J2000) &
(Arcmin) &
(Degrees) 
}

\startdata
960416 & ub & 10h09m15.65s & $+67\arcdeg07\arcmin36.8\arcsec$ & 23.9 & 25.706\nl
960727 & uk & 10h37m45.44s & $+42\arcdeg05\arcmin13.3\arcsec$ &  1.4 & 82.222\nl
961002 & uk & 23h30m18.78s & $-33\arcdeg07\arcmin44.4\arcsec$ &  2.1 & 81.847\nl
961019 & ub & 23h41m20.75s & $-31\arcdeg52\arcmin58.3\arcsec$ & 11.0 & 48.530\nl
970815 & ub & 10h38m36.27s & $+20\arcdeg33\arcmin43.9\arcsec$ &  2.8 & 68.519\nl
970828 & ub & 10h46m12.46s & $+19\arcdeg10\arcmin18.6\arcsec$ &  1.1 & 83.514\nl
971214 & ub & 11h32m40.33s & $+11\arcdeg03\arcmin06.3\arcsec$ & 13.2 & 54.331\nl
980703 & ub & 22h06m35.40s & $ -9\arcdeg03\arcmin55.7\arcsec$ & 13.5 & 33.132\nl
981220 & uk & 23h09m29.47s & $ +7\arcdeg20\arcmin04.3\arcsec$ &  0.8 & 67.142\nl
\enddata
\tablenotetext{a}{u - {\it Ulysses}; b - BATSE; k - KONUS}
\end{deluxetable}

GRB candidate 960529 (Fig.~\ref{tgrb}b) was detected in two SSCs over
two consecutive dwells, yielding four position determinations.  The
two smallest boxes are plotted in Figure~\ref{tgrb}b; the largest two
are consistent with the intersection of these two.  This event was
detected by KONUS, but not by {\it Ulysses}, rendering an IPN annulus
impossible to calculate.  Both the ASM and KONUS observed three
successive peaks, lasting a total of $\sim200$ s, but KONUS detected
no emission above 50~keV (P. Butterworth 1997, private communication).
It is therefore possible that this event was not a GRB but was a
series of hard flares from an unknown X-ray source.

GRB~960727 (Fig.~\ref{tgrb}c) and GRB~961002 (Fig.~\ref{tgrb}d) were
bright events, but both were seen in only one ASM SSC.  However, they
were both detected by KONUS and {\it Ulysses}, so annuli could be
calculated that reduced the lengths of the error boxes to $1.4\arcmin$
and $2.1\arcmin$, respectively.  GRB~961029 (Fig.~\ref{tgrb}f) was
seen as an abrupt rise in the last few seconds of a single-camera ASM
observation, so despite reaching a peak flux of $\sim120$~mCrab, the
ASM measured only a total of 289 counts from this GRB.  GRB~961230
(Fig.~\ref{tgrb}g) was also weak (see Table~\ref{wektab}), but was
detected in two cameras.  Neither GRB~961029 nor GRB~961230 was
detected by more than one IPN instrument, so triangulation annuli were
impossible to calculate in these cases.

Thirdly, as of May 1997, we began searching for GRB events in the
real-time ASM data.  We first developed software to respond directly
to the GRB alerts released by the BATSE team over the Gamma-ray burst
Coordinate Network (GCN).  The ASM observing plan cannot be changed in
response to alerts, but on occasion the ASM FOV will overlap with a
BATSE GRB error circle at or shortly after the trigger time.  GRBs
have been observed by BATSE to last hundreds of seconds, often with
multiple peaks (\markcite{meega96}Meegan et al. 1996), and there is
evidence that GRBs last longer at lower energies than at higher
energies (\markcite{fenim95}Fenimore et al. 1995;
\markcite{pirol98}Piro et al. 1998).  Since the ASM rotates every
90~s, a rotation may put the FOV of an SSC over a burst which is in
progress.  

We compare the BATSE information for each new trigger with the planned
observing schedule for the ASM and to alert us if the ASM is scheduled
to scan over the BATSE error circle within 1000 s from the time of
trigger.  This program has led to the detection of five GRBs as of
this writing, and the resulting ASM positions were disseminated to the
community within 2--12~h from the events, enabling rapid follow-up by
other observers.  The positions reported here represent refinements
that supercede any GRB positions previously reported in IAU or GCN
Circulars by the ASM team.  However, unless stated explicitly below,
these positions differ from the initial error boxes reported in the
Circulars by no more than an arcminute.

GRB~970815 (Fig.~\ref{tgrb}h) had multiple peaks in its light curve,
and it was located such that it was detected in a single-SSC during a
single 90-s dwell (\markcite{smith97}Smith et al. 1997).  During the
next dwell, it became much brighter while it was observed with both
SSCs~1 and 2.  As the event faded during a third dwell, the source
location was less than a degree from the edge of the FOV of SSC~1 (and
outside the FOV of SSC~2).  This location is outside the region of the
FOV considered in the present analysis, and a reliable position could
therefore not be obtained from this dwell.  As with GRB candidate
960529, only the two smallest error boxes are shown in
Figure~\ref{tgrb}h.  This GRB was also seen by both BATSE and {\it
Ulysses}, and an IPN annulus confirms the ASM position.  Three days
after the GRB event, a weak X-ray source with constant intensity near
the ASM position was seen by {\it ASCA} (\markcite{mura97a}Murakami et
al. 1997a, included in Fig.~\ref{tgrb}h); its relation to GRB~970815
remains uncertain.

GRB~970828 (Fig.~\ref{tgrb}i) was localized in two SSCs, and its
location was published within two hours of the onset of the event
(\markcite{remil97}Remillard et al. 1997).  The PCA on {\it RXTE}
slewed to observe the fading X-ray flux within 3.6~h, confirming the
ASM position (\markcite{marsh97}Marshall, Cannizzo, \& Corbet 1997).
This PCA observation remains the fastest capture of a GRB afterglow at
X-ray energies.

An IPN annulus based on BATSE and {\it Ulysses} detections was made
available $\sim1$~d after the ASM report (Hurley et al. 1997). A
fading X-ray source was observed with {\it ASCA} over the interval of
1.2 to 2.1~days after the burst trigger (\markcite{mura97b}Murakami et
al. 1997b). Later ROSAT observations narrowed the position to within a
radius of $10\arcsec$ (\markcite{grein97}Greiner et al. 1997).
Optical and radio instruments observed this location within 4~h of the
trigger time, but despite intense monitoring over the following weeks,
no counterpart was seen at wavelengths longer than X-rays
(\markcite{groot98}Groot et al. 1998).  

GRB~971024 (Figure~\ref{tgrb}j) was detected in two cameras, but the
event proved to be extremely weak in the ASM energy band.  Due to its
position near the edge of the field of view of SSC~2, only 159~counts
were detected.  The position analysis described above indicates that
positions for detections this weak are unreliable, so we do not report
an error box from this SSC.  The error box derived from the detection
of GRB~971024 in SSC~1 is shown in Figure~\ref{tgrb}j.  Because of the
extremely large error region, very little follow-up was performed.  No
candidate counterpart was reported.

GRB~971214 (Fig.~\ref{tgrb}k) was detected by SSC~3 within a single
dwell.  We reported a line of position via the GCN about 2.3~h after
the event.  This GRB was also detected with BATSE and {\it Ulysses},
and the resulting IPN annulus, reported two days later, was
$7.9\arcmin$ in width (\markcite{kippe97}Kippen et al. 1997).  The
error box reported here is smaller in area than the initial ASM
position by about a factor of three.  This burst was also detected
simultaneously and independently in the Wide-Field Camera of {\it
BeppoSAX}, generating a 99\% confidence error circle $3.9\arcmin$ in
radius (\markcite{heise97}Heise et al. 1997).  This region was further
reduced through pointings by the Narrow-Field Instruments (NFI), which
localized a fading X-ray counterpart to $1\arcmin$
(\markcite{anton97}Antonelli et al. 1997).  The two error circles are
shown in Figure~\ref{tgrb}k.  A fading optical source was quickly
identified (\markcite{halpe97}Halpern et al. 1997), and later
observations indicated that the host galaxy candidate had a redshift
of $z=3.42$ (\markcite{kulk98a}Kulkarni et al. 1998a).

GRB~980703 (Figure~\ref{tgrb}l) was a bright burst, the onset of which
was seen in two SSCs simultaneously.  The ASM position was first
reported $\sim12$~h after the event (\markcite{levin98}Levine, Morgan,
\& Muno 1998).  Observations of a region centered on the ASM
localization with the {\it BeppoSAX} NFI at 22~h after the BATSE
trigger were able to localize a fading X-ray source to within a circle
of $50\arcsec$ radius (\markcite{gala98a}Galama et al. 1998a).  An IPN
annulus $18.4\arcmin$ in width was made available within 3~d
(\markcite{hurle98}Hurley \& Kouveliotou 1998).  The final IPN annulus
reported here is $13.5\arcmin$ wide; it is consistent with the ASM and
{\it BeppoSAX} measurements.  Radio and optical observers were also
able to identify counterparts (\markcite{frail98}Frail et al. 1998).
Later spectroscopy revealed a redshift of $z=0.9653\pm0.0007$
(\markcite{djorg98}Djorgovski et al. 1998), the third cosmological
redshift to be measured for a GRB source.

\end{multicols}
\refstepcounter{figure}
\PSbox{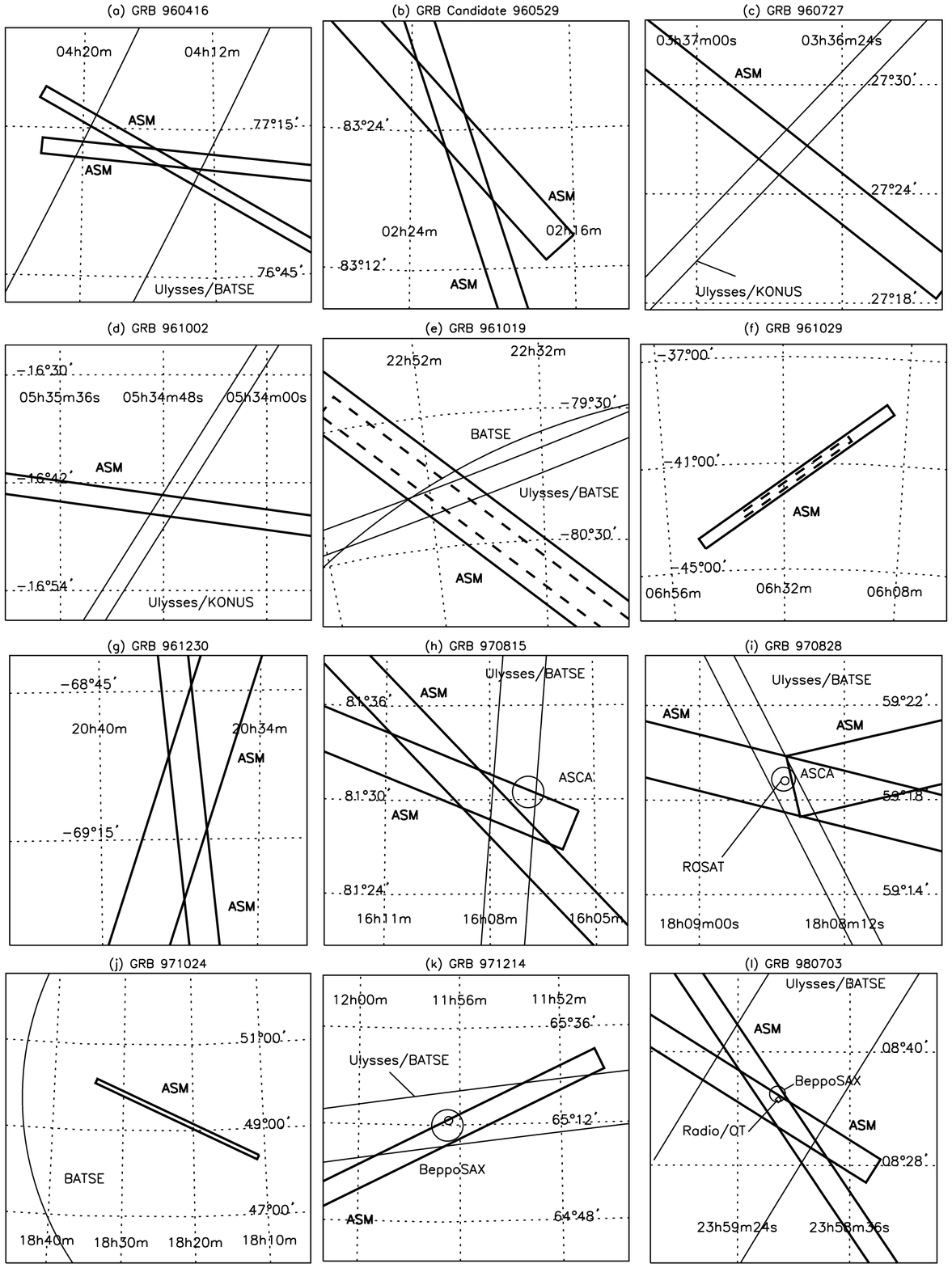 hoffset=20 voffset=0 hscale=92 
vscale=92}{5.0in}{8.9in} {\\\small Fig. 6 -- The ASM localizations of
12 gamma-ray bursts, shown with position information from other
satellites when available.  Each frame is mapped at a different scale.
ASM (dark lines) and BATSE locations are given at 90\% confidence,
while IPN annuli are $3\sigma$.  In cases of dim GRBs, boxes at 68\%
confidence are also plotted as dashed outlines.  Positions and sizes
of the 90\% regions are given in
Tables~\protect{\ref{flutab}}--\protect{\ref{cortab}}, while
references to the other instruments are given in the text and in
Tables~\protect{\ref{othtab}}--\protect{\ref{ipntab}}.\label{tgrb}}
\section*{}
\vskip -15pt

\refstepcounter{figure}
\PSbox{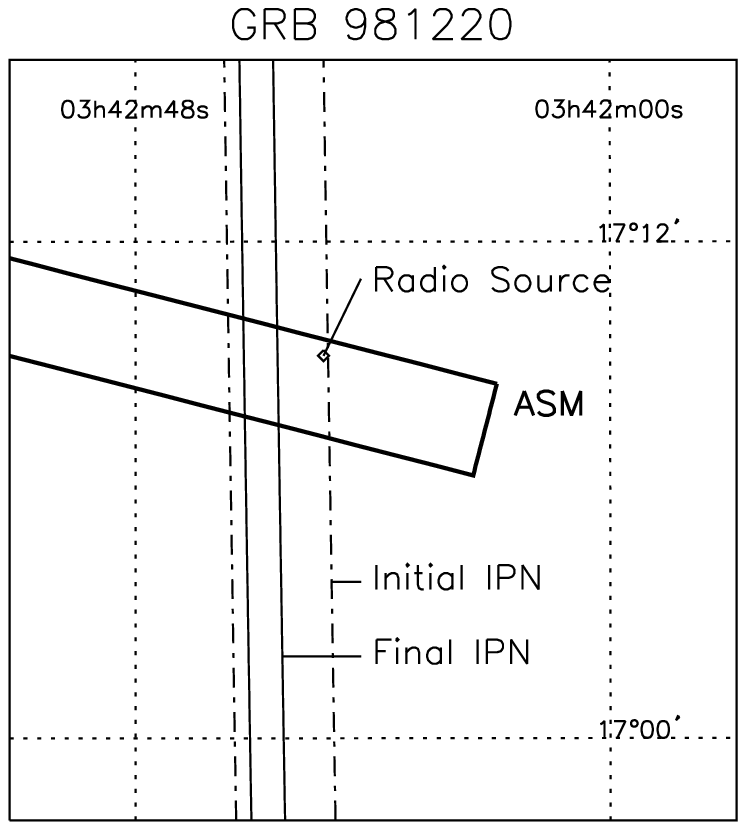 hoffset=-15 voffset=0 hscale=102 vscale=102}{3.5in}{3.5in} 
{\\\small Fig. 7 -- The ASM localization of GRB~981220, shown with two
IPN annuli constructed by triangulation of burst arrival times at {\it
Ulysses} and KONUS/{\it WIND}.  The radio source at
$\alpha$~=~03h42m28.98s$\pm0.07$s, $\delta =
+17\arcdeg09\arcmin14.7\arcsec\pm1.6\arcsec$ (J2000) was discovered by
\protect{\markcite{gala98b}}Galama et al. (1998) and found to be
highly variable by \protect{\markcite{fraku98}}Frail \& Kulkarni
(1998).  The final IPN annulus reported here (see
Table~\protect{\ref{ipntab}}) excludes this radio source as a
counterpart to GRB~981220, and further monitoring with the VLA shows
behavior inconsistent with that of other GRB afterglow at radio
wavelengths (\protect{\markcite{frkut99}}Frail, Kulkarni, \& Taylor
1999).\label{decgrb}}

\vspace{0.4cm}

Fourthly, in September of 1997, we established a ``self-trigger''
system, in which the incoming time-series data is checked for excess
variability using the same criteria developed during the archival
search described above.  Dwells with significant non-linear time
series sequences are flagged as possible GRB events.  If the standard
cross-correlation analysis also reports a possible new source
detection at better than 5~$\sigma$, an email alert is distributed.
If BATSE and/or {\it BeppoSAX} data indicate that a GRB was active
during the time of the dwell, the significance limit is lowered to
3~$\sigma$, to ensure that a dim GRB does not slip through the system.
This system has detected one GRB to date that was not also detected by
BATSE.

GRB~981220 (Figure~\ref{decgrb}) was detected by SSC~2 during a single
dwell, and an alert was distributed 32~h after the initial event.  An
IPN annulus $2.4\arcmin$ wide using {\it Ulysses} and KONUS detections
was rapidly calculated (\markcite{hural98}Hurley et al. 1998).
Although no optical transients were reported, a rapidly varying radio
source was discovered within the ASM/IPN error box and attributed to
GRB~981220 (\markcite{gala98b}Galama et al. 1998b;
\markcite{fraku98}Frail \& Kulkarni 1998).  The final IPN annulus
reported in Table~\ref{ipntab} is $0.8\arcmin$ wide and excludes this
radio source as a counterpart to GRB~981220.  Further monitoring with
the VLA has shown that the behavior of this source's light curve does
not resemble that of other GRB afterglows (\markcite{frkut99}Frail,
Kulkarni, \& Taylor 1999).  This radio source is therefore most likely
unrelated to GRB~981220.  The nature and characteristics of any
afterglow from GRB~981220 remain unknown at this time.

Seven detections of five of these thirteen bursts yielded numbers of
counts low enough for the chance that each of the derived positions is
spurious to be $\gtrsim7$\%.  The probabilities of spurious detections
in each coordinate as predicted by Equation~\ref{spur} are given in
Table~\ref{wektab} for all ASM GRB detections.  Three weak bursts,
GRB~961019 (Fig.~\ref{tgrb}e), GRB~971024 (Fig.~\ref{tgrb}j), and
GRB~971214 (Fig.~\ref{tgrb}k), have positions derived from BATSE, IPN,
or {\it BeppoSAX} which confirm the ASM positions.  In the case of
GRB~961019 (Fig.~\ref{tgrb}e), the IPN annulus significantly reduces
the size of the error box.  In the case of GRB~961230
(Fig.~\ref{tgrb}g), the weak ASM positions confirm each other (the
probability that two spurious boxes overlap by chance is less than
$10^{-3}$), but there are no independently derived positions to
compare with the ASM error region for GRB~961029 (Fig.~\ref{tgrb}f).

Three further GRBs (GRB~961216, GRB~971216 and GRB~981005) had BATSE
positions that were consistent with the ASM FOV at the time of
trigger, and ASM analysis indicated the presence of uncatalogued X-ray
source candidates in the FOV of at least one SSC.  In the case of
GRB~961216, a $\sim4\ \sigma$ peak was detected less than a degree from
the edge of the FOV of a single SSC.  This position lies outside the
region of the FOV included in the present analysis, so we do not
report it here.  Observations at the times of GRB~971216 and
GRB~981005 were more complex.  The highest peaks in the
cross-correlation maps derived from SSCs~1 and 2 had low significance
($\sim$2--4~$\sigma$), fell below the 200-count lower limit for
reliability, and mapped to inconsistent celestial locations.  We are
therefore unable to report reliable detections for either of these
bursts.  Furthermore, we cannot provide useful upper limits for the
X-ray fluxes from these GRBs, since it is possible that the actual GRB
sources were located outside the FOV of the ASM during all of these
observations.

Although we know of no other bright, burst-like events in the ASM
database that we cannot identify, it is possible that the ASM has
detected GRBs other than the ones reported here.  Our understanding of
how to distinguish real short-lived X-ray events from solar- or
particle-induced events has improved since the archival search
described above was completed.  The difficulty in identifying real
events in the archival search led us to exclude approximately
one-third of the data from consideration.  We may also miss GRBs in
the real-time search, if the telemetry stream from the satellite is
interrupted by internet or server outages on the ground, or if the
packets are received out of order.  These problems are corrected in
production data, which are available $\sim$1--4 days after the
observations, but these data are not currently being searched for
GRBs.  Projects to reprocess the archival data and utilize the
production data to create a complete ASM GRB catalog are planned.

We argue above that we have identified nearly all the GRBs above a
certain threshhold in the $\sim2\times10^7$~s~sr of sky coverage
examined via the searches reported here (roughly 60\% of the first
three years of ASM observations).  The observed burst rate is
therefore $7\times10^{-7}$~s$^{-1}$~sr$^{-1}$.  The BATSE 4B catalog
records a full-sky burst rate of $1.7\times10^{-6}$~s$^{-1}$~sr$^{-1}$
when corrected for the BATSE sky exposure (Paciesas et
al. \markcite{pac4B}1999), although this rate ignores non-triggered
GRBs such as those found by Kommers et al.  (\markcite{klkpp97}1997).
If the ASM can detect all the BATSE GRBs that appear in its FOV, we
would expect to find 33~GRBs in our subset of the ASM database.

\begin{deluxetable}{ccccc}
\tablecaption{Probability of Misidentifying Weak Bursts\label{wektab}}
\tablehead{
Date of GRB &
SSC &
Number &
Chance of spurious &
Chance of spurious \nl
(yymmdd) & 
Number &
of counts &
detection in $\phi$ (\%) & 
detection in $\theta$ (\%) 
}

\startdata
960416 & 1 &  997 &  2 &  3 \nl
960416 & 2 & 1424 &  4 &  2 \nl
960529 & 1 & 1035 &  1 &  3 \nl
960529 & 1 & 1043 &  1 &  3 \nl
960529 & 2 & 2223 &  3 &  2 \nl
960727 & 2 & 1899 &  3 &  2 \nl
961002 & 2 & 1498 &  4 &  2 \nl
961019 & 2 &  297 & 23 & 38 \nl
961029 & 2 &  289 & 24 & 41 \nl
961230 & 1 &  329 &  8 & 12 \nl
961230 & 2 &  386 & 14 & 19 \nl
970815 & 2 & 1060 &  4 &  3 \nl
970815 & 1 & 4843 &  1 &  2 \nl
970815 & 2 & 3769 &  3 &  1 \nl
970828 & 1 & 1348 &  1 &  3 \nl
970828 & 2 & 1230 &  4 &  3 \nl
971024 & 1 &  303 & 10 & 14 \nl
971214 & 3 &  662 &  6 & 13 \nl
980703 & 1 & 1539 &  1 &  3 \nl
980703 & 2 & 1622 &  3 &  2 \nl
981220 & 2 & 2236 &  3 &  2 \nl
\enddata
\end{deluxetable}

Clearly, the simple assumptions of this calculation of the rate of
ASM-detectable GRBs are inadequate.  In section~\ref{sec:mod}, we show
that the ASM is less sensitive to a source closer to the edge of the
FOV than to a source of identical brightness at the center of the FOV.
The calculation of the ASM sky coverage is therefore overestimated.
Also, it is not clear how the characteristics of the BATSE GRB
population extrapolate to X-ray energies.  All of the bursts reported
here lasted more than ten seconds, while the BATSE duration
distribution includes a significant population below 1~s (Paciesas et
al. \markcite{pac4B}1999).  Although the ASM has detected short bursts
from Soft Gamma Repeaters (Smith, Bradt, \& Levine
\markcite{sbl99}1999), we have found no other short, bright events
that might be GRBs.  A thorough treatment of the burst rate is beyond
the scope of this paper, and we defer it to a later investigation.

\section{Summary}

This paper describes an empirical method to characterize the error
distribution of the ASM source-localization analysis.  We used
$\sim14,000$ observations of sources at random positions in the
central $9.2\arcdeg \times 90\arcdeg$ region of the FOV.  An ASM error
box derived from a single SSC detection takes the shape of a long,
thin rectangle.  We found that the accuracy (at 95\% confidence) of
the derived positions in the short direction ($\phi$) is fairly steady
at $\pm1.2\arcmin$ ($\pm1.9\arcmin$ in SSC~3) for bright sources, but
increases rapidly as the detected number of counts falls below
$\sim700$.  The error in the long direction ($\theta$) does not level
off as dramatically for bright sources, but decreases with increasing
source brightness, ranging from about $\pm1.5\arcdeg$ to about
$\pm12\arcmin$ for observations yielding $\gtrsim700$~counts.  The
chance that an error box represents a spurious detection rises rapidly
for detections yielding $\lesssim700$~counts, and detections with less
than 200~counts (300~counts in SSC~3) do not yield reliable positions.

We apply this method to the localization of GRBs detected
serendipitously by the ASM.  Through realtime monitoring and archival
searches, we have localized thirteen GRBs detected in the first three
years of ASM observations.  This list is most likely not complete,
because it has not yet been possible to thoroughly search the entire
database of ASM production data.  The ASM is not designed to
distinguish between GRBs and X-ray bursts; we have relied on
confirmations from other instruments to identify these events.
Nevertheless, we found no unconfirmed burst-like events that we could
not identify with known X-ray sources.  Of the thirteen GRBs presented
here, six were observed in two SSCs, so the GRBs were localized within
diamond-shaped error boxes a few arcminutes on a side.  The error
boxes from each of five further single-SSC detections were combined
with an IPN annulus to obtain joint error boxes of a few arcminutes in
each of two dimensions.  These error boxes are useful for follow-up
studies of GRBs at other wavelengths.

Rapid analysis of ASM data has led to several successful and
provocative discoveries.  The initial ASM position of GRB~970828 was
released within 2~h of the burst event, enabling X-ray, optical and
radio observations to be performed within 3.6~h of the initial
trigger.  While a fading X-ray counterpart was detected with {\it
ASCA} and {\it ROSAT}, no counterparts were seen at longer wavelengths
to a limiting magnitude of $\sim24$ (\markcite{groot98}Groot et
al. 1998).  The thorough coverage in optical and radio bands of this
well-constrained area established convincingly that GRB afterglows
span a wide range of intensities.  Groot et
al. \markcite{groot98}(1998) suggest two possible explanations for the
faintness of any optical counterpart: beaming in the blast wave could
be directing the emission away from the Earth, or strong extinction
enhanced by the unknown redshift, could be reducing the apparent
magnitude of the source below detectable limits. This latter
hypothesis might imply that GRBs occur in star-forming regions, as
predicted by the ``hypernovae'' class of GRB models
(\markcite{paczy98}Paczy\'{n}ski 1998).

In the case of GRB~980703, the ASM position was reported within 12~h
of the burst event (\markcite{levin98}Levine et al. 1998), which led
to the rapid identification of a fading optical counterpart as well as
a rapidly varying radio counterpart (\markcite{frail98}Frail et
al. 1998).  Further spectroscopic observations of the fading optical
transient led to the measurement of a redshift of $z=0.9653 \pm
0.0007$ (\markcite{djorg98}Djorgovski et al. 1998); the third
cosmological redshift ever measured for a GRB.

\acknowledgments

The authors wish to thank Saul Rappaport and John Doty for helpful
discussions, as well as Scott Barthelmy for his inestimable service in
providing the GCN.  KH is grateful for support under JPL Contract
958056 for Ulysses operations, and to NASA Grant NAG~5~3811 for
support of the IPN.  Support for this work was provided in part by
NASA Contract NAS5--30612.

\end{document}